\documentclass{an}
\usepackage{graphicx}
\usepackage{times}
\newcommand{\pun}[1]{\mbox{\rm\,#1}} 

\newcommand{\logg}{\ensuremath{\log g}}

\newcommand{\Teff}{\ensuremath{T_{\mathrm{eff}}}}

\newcommand{\ReM}{\ensuremath{\mathrm{Re}_\mathrm{M}}}

\newcommand{\beq}{\begin{equation}}
\newcommand{\eeq}{\end{equation}}

\Volume{000}             
\Year{0000}              
\Month{00}               
\Pagespan{000}{000}      

\begin{document}
\lhead[\thepage]{Dorch \&\ Ludwig: Small-scale magnetic fields on late-type 
 M-dwarfs}
\rhead[Astron. Nachr./AN~{\bf XXX} (200X) X]{\thepage}
\headnote{Astron. Nachr./AN {\bf 32X} (200X) X, XXX--XXX}

\title{Small-scale magnetic fields on late-type M-dwarfs}

\author{S.B.F. Dorch$^{1,2}$ \and\ H.-G. Ludwig$^3$}
\institute{%
The Niels Bohr Institute for Astronomy, Physics and Geophysics,
Juliane Maries Vej 30, DK-2100 Copenhagen {\O}, Denmark
\and
The Institute for Solar Physics of the Royal Swedish Academy of Sciences,
SCFAB, SE-10691 Stockholm, Sweden
\and
Lund Observatory, Box 43, S-22100 Lund, Sweden} 

\date{Received {\it date will be inserted by the editor};
accepted {\it date will be inserted by the editor}}

\abstract{We performed kinematic studies of the evolution of
small-scale magnetic fields in the surface layers of M-dwarfs.  We
solved the induction equation for a prescribed velocity field,
magnetic Reynolds number \ReM, and boundary conditions in a Cartesian
box, representing a volume comprising the optically thin stellar
atmosphere and the uppermost part of the optically thick convective
envelope.  The velocity field is spatially and temporally variable,
and stems from detailed radiation-hydrodynamics simulations of
convective flows in a proto-typical late-type M-dwarf
($\Teff=2800\pun{K}, \logg=5.0$, solar chemical composition, spectral
type $\approx$M6).  We find dynamo action for $\ReM \ge 400$. Growth
time scales of the magnetic field are comparable to the convective
turn-over time scale ($\approx 150\pun{sec}$). The convective velocity
field concentrates the magnetic field in sheets and tubular structures
in the inter-granular downflows.  Scaling from solar conditions
suggests that field strengths as high as 20\pun{kG} might be reached
locally.  Perhaps surprisingly, \ReM\ is of order unity in the surface
layers of cooler M-dwarfs, rendering the dynamo inoperative.  In all
studied cases we find a rather low spatial filling factor of the
magnetic field.  } \correspondence{dorch@astro.ku.dk}

\maketitle

\section{Introduction}

M-type dwarfs show the highest degree of magnetic activity.  Besides
X-ray and $\mathrm{H}_\alpha$ emission as indirect tracers, Zeeman
broadening of magnetically sensitive photospheric lines has been
observed (Johns-Krull \&\ Valenti 1996, Kochukhov et al. 2001). The
photospheric lines show no rotational modulation or net polarization,
indicating that fields of {\em small scale\/} (relative to the star's
radius) exist on the surfaces of M-dwarfs. These fields can have an
appreciable strength (few\pun{kG}), and can cover a substantial area
(filling factors up to 50\pun{\%}).

Little is known about the structure of magnetic fields in the
photospheres of M-dwarfs. From an observational point of view one
would like to get some input from theory, that would alleviate the problem of 
disentangling field strength and filling factor. To this end we
started to investigate the kinematic effect of the convective velocity
field on a magnetic seed field. This has become possible due to recent
progress in the hydrodynamical modeling of atmospheres in the regime
of cool M-dwarfs (Ludwig, Allard, \&\ Hauschildt 2002).  The present
investigation is targeting at an application to spectroscopy, meaning
that we focus on layers near optical depth unity.  To overcome
partially the limits stemming from the restriction to the kinematic
magneto-hydrodynamics regime (hereafter ``kinematic MHD''), we tried
to relate our results to the situation in the Sun, emphasizing the
scaling from familiar solar conditions to M-dwarfs.

\section{Model setup}

\subsection{The convective velocity}

\begin{figure}
\resizebox{\hsize}{!}
{\includegraphics[]{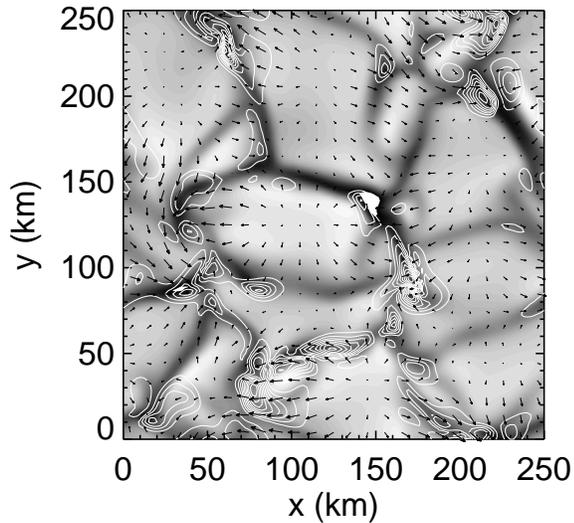}}
\caption{Snapshot of the vertical component of the magnetic field at optical depth unity ({\bf contours}), emergent intensity ({\bf grey tones}), and the velocity field ({\bf arrows}) towards the end of an evolutionary sequence of an unquenched model with \mbox{\ReM=20}.}
\label{fig-convbz}
\end{figure}

The velocity field was taken from a radiation-hydrodynamics
simulation of a prototypical M-dwarf atmosphere (Ludwig et a. 2002,
see Fig~\ref{fig-convbz}) of $\Teff=2800\pun{K}$, $\logg=5.0$, and
solar chemical composition (corresponding to a spectral type of
$\approx$M6). The 3D radiation-hydrodynamics code of Nordlund \&\
Stein (Stein \&\ Nordlund 1998, and references therein) has been
adapted for this purpose. The code treats the interaction of radiation
and gas flows in detail, rotation and magnetic fields are
neglected. 
Rotation has only a minor effect on the structure of the convective
flows near the stellar surface due to the short convective turn-over
time scales (relative to typical rotation periods of several hours or
longer). Magnetic fields can potentially influence the flow structure
significantly provided the field is strong, i.e. dynamically
important. However, this corresponds to the full MHD case which is
beyond the scope of this paper.
We selected a temporal sequence towards the end of the
simulation run, which comprises 1500\pun{s} of stellar time,
corresponding to roughly 10 convective turn-over time scales. The
sequence consists of 150 snapshots of the flow field, each comprising
\mbox{$125\times 125 \times 82$} grid points corresponding to
\mbox{$250\times 250\times 87\pun{km}^3$} in geometrical size. At any
instant in time about 10 granular cells were present in the
computational domain, ensuring a statistically representative
ensemble. M-dwarf granulation is on the qualitative level similar to
the solar granulation. Quantitatively, there are differences which are
discussed below where relevant.

\subsection{Kinematic magneto-hydrodynamics}

We assume the kinematic regime of MHD, where one
neglects the back-reaction of the magnetic field on the fluid
motions. 
This means that the non-linear behavior cannot be rendered correctly in 
this regime, which is also sometimes referred to as the linear regime
of MHD.
Then solving the MHD equations reduces to the problem of
seeking the solution to the time-dependent induction equation:
\beq
 \frac{\partial {\bf B}}{\partial t} = \nabla\times ({\bf u}\times {\bf B})
 + \eta \nabla^2 {\bf B}, \label{induction.eq}
\eeq
where ${\bf u}$ is the prescribed velocity field (from the
radiation-hydrodynamic simulation), ${\bf B}$ the magnetic field, and
$\eta$ the magnetic diffusivity. We assume a spatially constant
magnetic diffusivity. Note, that the problem is linear on this level
of approximation. We vary the magnetic diffusion by setting the
magnetic Reynolds number \ReM\ which is defined as
$\mathrm{Re}_\mathrm{M} = {\rm U} \ell/\eta$, where U is a
characteristic velocity, $\ell$ a characteristic length scale, and
$\eta$ the magnetic diffusivity, which in turn is related to the
electrical conductivity $\sigma$ through $\eta = 1/\mu \sigma$.

We solve Eq.\ (\ref{induction.eq}) using staggered variables on the
grid of the hydrodynamical flow field. The numerical method was
originally developed by Galsgaard and others (Galsgaard \& Nordlund
\cite{Galsgaard+Nordlund97}) for general MHD purposes. A special
version is the code used by Archontis \& Dorch (Dorch
\cite{Dorch2000}; Archontis, Dorch, \&~Nordlund \cite{Archontis+ea02})
to study dynamo action in prescribed flows.

We kept the same velocity field, and studied different magnetic
diffusivities. To first order we hope to represent the different
conditions encountered in various M-dwarf atmospheres by the the
change of magnetic diffusivities.

\subsection{Magnetic quenching of the velocity field}

The kinematic approximation to the MHD equations --- i.e.\ neglecting
the back-reaction of the magnetic field by omitting the Lorentz force
in the equation of motion ---
becomes inaccurate when the magnetic field becomes sufficiently
strong. That is, when the magnetic energy density approaches
equipartition with the kinetic energy density, corresponding to a
field strength ${\rm B}_{\rm eq} = {\rm u} \sqrt{\mu \rho}$.
One can try to capture the effect of a near-equipartition
magnetic field, by utilizing a quenched velocity field that has a
reduced amplitude in the regions where the magnetic field is
strong. In a few cases we employed such a quenched velocity field when
solving the induction equation Eq.\ (\ref{induction.eq}).  The
quenched velocity field ${\bf u}_{\rm q}$ was modeled according to
\beq
 {\bf u}_{\rm q}={\bf u}~ \exp {-\alpha (e_{\rm M}/e_{\rm K})^2}, 
  \label{quench.eq}
\eeq
where $\alpha$ is a constant, $e_{\rm M}$ the local magnetic energy
density, $e_{\rm K}$ the average kinetic energy density, taken to be
constant in time and space. 
The quenching reduces the velocity amplitude at 
locations where the magnetic energy (or field) becomes comparable to
the kinetic energy of the fluid motions: it thus reduces the growth of
the magnetic field in these regions causing it to saturate.
We chose $\alpha = 1.75$
which ensures ${\bf u}_{\rm q} \approx {\bf u}$ up to field strength
of 10\% of ${\rm B}_{\rm eq}$, and ${\bf u}_{\rm q} \approx 0$ at a
field strength of 1.2 times equipartition.  Note, that this quenching
procedure makes the problem non-linear in B, and that a constant 
$e_{\rm K}$ merely introduces a scaling factor in the
solution, i.e. its concrete value is unimportant.  In future work we
plan to replace Eq.\ (\ref{quench.eq}) with a more realistic quenching
expression, taking into account the geometry of the flow and the
magnetic field.

\subsection{Initial and boundary conditions}

The numerical method used to solve Eq.\ (\ref{induction.eq}) by
default considers periodic boundary conditions in all three
dimensions. We implement non-periodic vertical boundary conditions
through ``ghost zones'' at the top and bottom of the domain: then one
may constrain the magnetic field to be vertical (horizontal) by
requiring its horizontal (vertical) component(s) to be anti-symmetric
across the boundary, and the remaining component(s) to be symmetric.

The initial magnetic field is either uni-directional (vertical or
horizontal) or smoothly varying on small scales.  The initial field
strength ${\rm B}_0$ is set to be much smaller than ${\rm B}_{\rm
eq}$ (only relevant for the cases with quenched velocity field).

\section{Results}

\subsection{Magnetic Reynolds numbers of order unity in cooler M-dwarfs}

\begin{figure*}[t!]
\resizebox{0.5\hsize}{!}
{\includegraphics[]{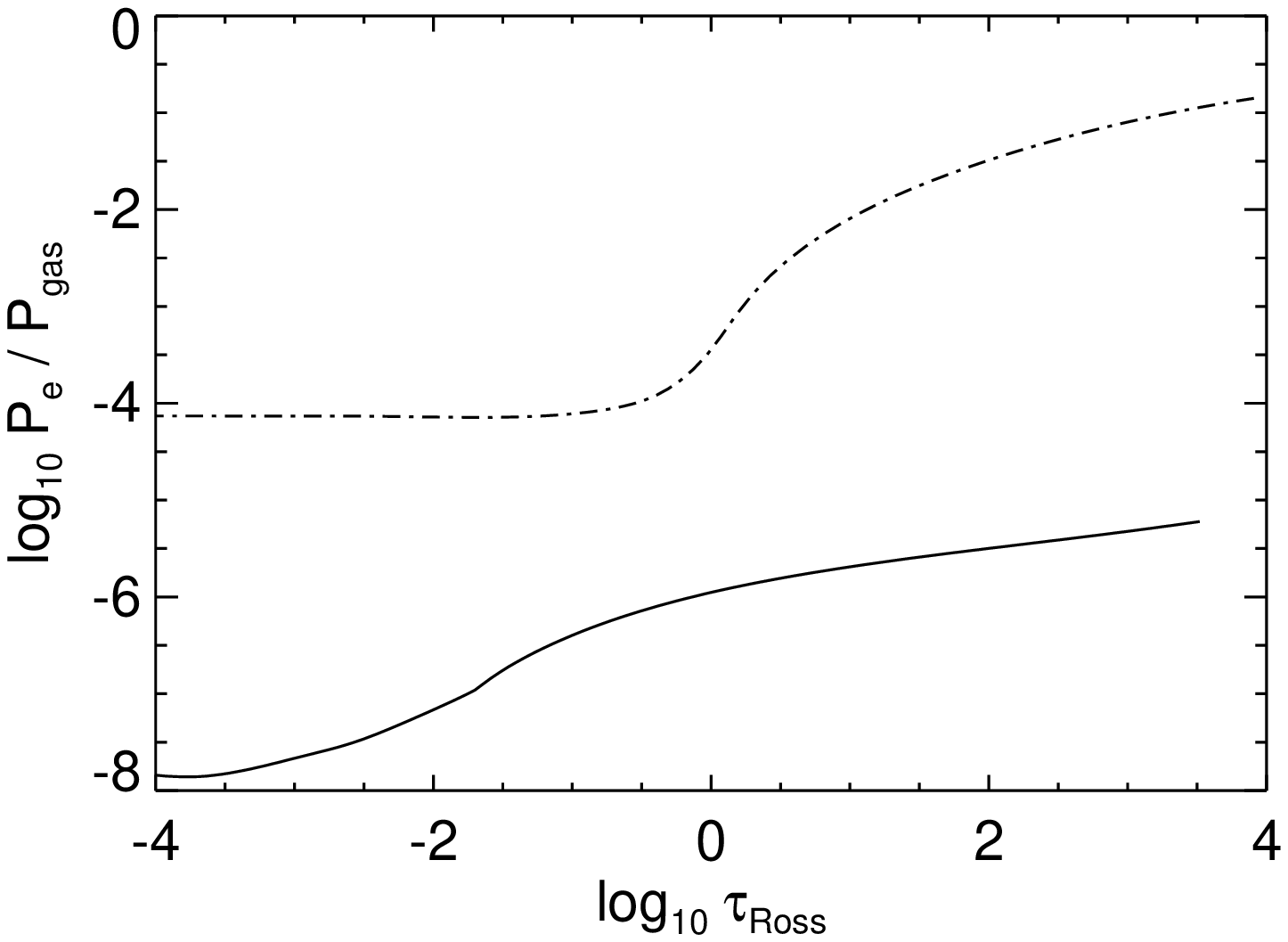}}
\resizebox{0.5\hsize}{!}
{\includegraphics[]{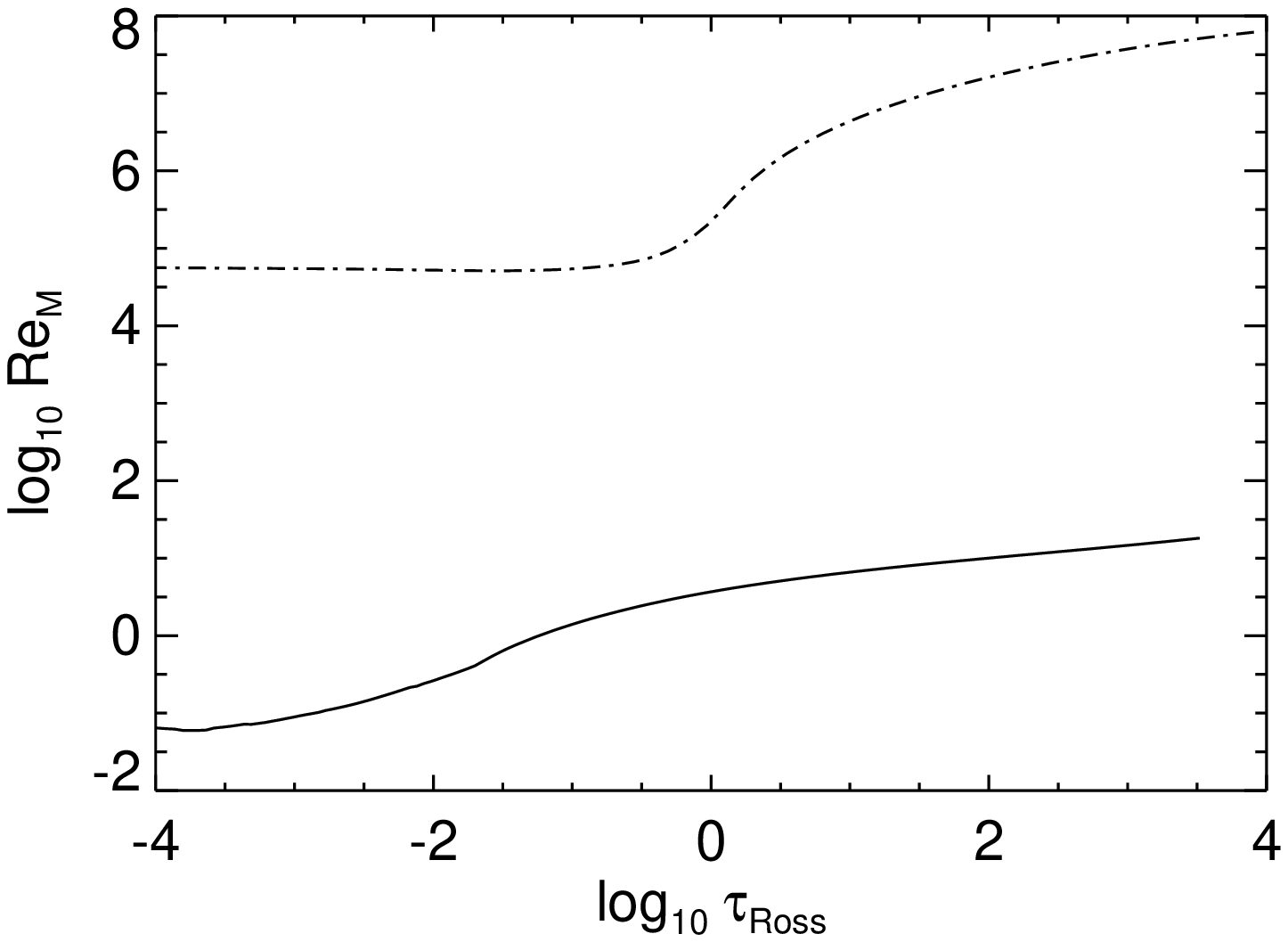}}
\caption[]{The ratio of electron to gas pressure ({\bf left panel}),
and the magnetic Reynolds number ({\bf right panel}) as a function of
Rosseland optical depth in a \mbox{$\Teff=2800\pun{K}$} M-dwarf ({\bf
solid line}) and a solar model atmosphere ({\bf dashed line}). Note
the low magnetic Reynolds number in the M-dwarf model, primarily
reflecting the low electric conductivity of the stellar gas in the
rather cool M-dwarf atmosphere.}
\label{fig-pepgas}
\end{figure*}

In the following we discuss kinematic MHD models with \ReM\ as low as
20. In the astrophysical context, low values of \ReM\ are uncommon,
mostly due to the large spatial scales usually involved. However, in
M-dwarf atmospheres we can be confronted with the situation of rather
low \ReM\ (Meyer \&\ Meyer-Hofmeister 1999): Fig.~\ref{fig-pepgas}
shows the ratio of electron to gas pressure, and \ReM\ in the
\mbox{$\Teff=2800\pun{K}$} M-dwarf model as well as the Sun. \ReM\ has
been evaluated assuming constant typical length and velocity scales
independent of depth\footnote{Sun: $\ell=1500\pun{km}$,
$v=2.4\pun{km/s}$; Mdwarf: $\ell=80\pun{km}$,
$v=0.16\pun{km/s}$}. \ReM\ primarily reflects the run of the electric
conductivity in the atmosphere, which in turn is mostly controlled by
the electron to gas pressure. The conductivity has been evaluated
assuming a weakly ionized plasma (Kopeck{\`y} 1957, as quoted by Stix
1989).

Even if one concedes an appreciable degree of uncertainty related to
the choice of scales, at sufficiently cool temperatures \ReM\ reaches
order unity in the surface layers. This is a consequence of the
declining electron density, the shrinking of spatial scales (due to
smaller pressure scale heights), and smaller convective velocities
(due to lower energy fluxes and higher atmospheric densities). This
refers to the surface layers. Qualitatively, we expect a strong
increase of \ReM\ with depth, and beyond a certain depth the regime of
$\ReM>1$ is reached again.  However, gas motions in this depth will
generally be slower and the tangling of magnetic field lines less rapid,
which may reduce the efficiency of chromospheric and coronal
heating. Whether this plays a role for the observed decline of stellar
activity at the transition from M- to L-dwarfs (Gizis et al. 2000) is
presently a matter of debate (Mohanty et al. 2002, Berger 2002).

\subsection{Scaling of the field strength}

\begin{figure}
\resizebox{\hsize}{!}
{\includegraphics[]{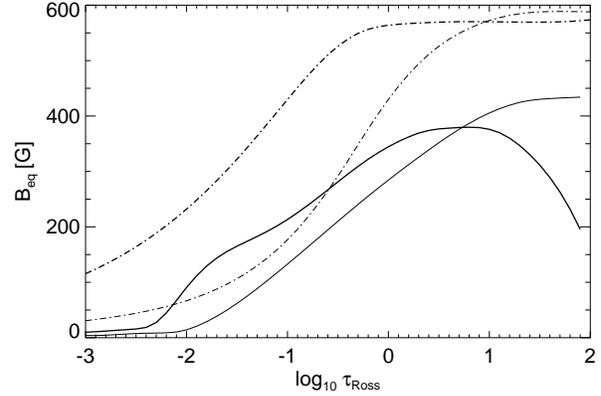}}
\caption{%
Kinetic energy densities expressed in terms of the corresponding
equipartition field strength as a function of Rosseland optical
depth. Depicted are results from hydrodynamical simulations for the
\mbox{$\Teff=2800\pun{K}$} M-dwarf ({\bf solid lines}) and the Sun
({\bf dash-dotted lines}). The total kinetic energy density is broken
into contribution of horizontal ({\bf thick lines}) and vertical
motions ({\bf thin lines}).}
\label{fig-beq}
\end{figure}

In order to obtain an estimate of the field strength to be expected in
M-dwarf atmospheres --- provided local conditions indeed control its
value --- we compared kinetic energy densities and pressures obtained
from our M-dwarf radiation-hydrodynamics model with a similar model
computed for the Sun. Figure~\ref{fig-beq} shows that the kinetic
energy densities in the surface layer of our reference M-dwarf model
are systematically smaller than in the Sun, but only within a factor
of~2. At the $\tau=1$ level the pressure is higher by a factor of~12
and density by a factor of~42 in the M-dwarf model relative to a solar
model (not shown).  Assuming a scaling of the magnetic field with gas
pressure, we expect a maximum field strength in flux concentrations of
about 10\pun{times} the solar value, or about 20\pun{kG} if one scales
values found in magneto-hydrodynamical simulations for the Sun (Stein
et al. 1999). However, the kinetic energy densities point towards a
field of slightly sub-solar field strengths.

\subsection{Dynamo action and magnetic Reynolds number}

\begin{figure}
\resizebox{\hsize}{!}
{\includegraphics[]{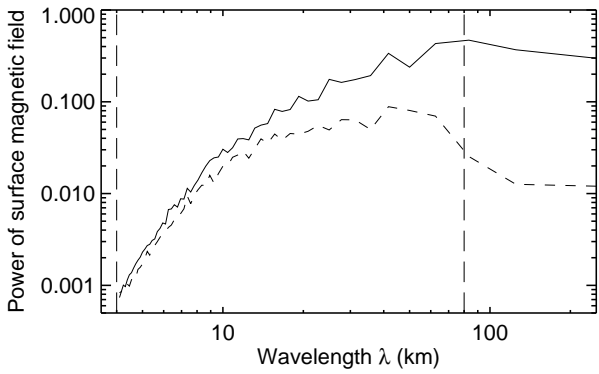}}
\caption{The power of the magnetic field at the surface
($\tau = 1$) in a model with \ReM = 100. Shown are the total field
strength ({\bf full curve}) and the vertical field component 
({\bf dashed}).  The vertical lines indicate the Nyquist wavelength 
({\bf left}) and the typical size of the granular cells ({\bf right}).}
\label{fig-hpower}
\end{figure}

A flow is a fast (kinematic) dynamo when the exponential growth
rate $\gamma$ is positive, where
\beq
 \gamma = \lim_{\eta \rightarrow 0} \gamma_{\eta} = \lim_{\eta \rightarrow 0}
 \lim_{t \rightarrow \infty}
 \log ({\rm E}_{\rm M}(t)/{\rm E}_{\rm M}(0))/t,
\eeq 
where ${\rm E}_{\rm M}$ is the total magnetic energy, and
$\gamma_{\eta}$ is the growth rate at a specific diffusivity,
corresponding to a certain \ReM. That is, fast  dynamo action
requires a continuous increase of magnetic energy, even in the limit
of vanishing diffusivity. This limit is relevant because most
astrophysical systems have $\ReM \gg 1$ and small but non-zero
$\eta$. It is believed (but not proven) that turbulent
astrophysical systems are fast dynamos operating at $\ReM \gg 1$.
When \ReM\ increases, the length scale of magnetic islands
decreases and scales as $\ReM^{-\frac{1}{2}}$. There is a maximum
value of \ReM, which can be achieved with our
numerical resolution $\Delta x = 2$ km, and therefore the largest
magnetic Reynolds number that we can allow is of the order of 400
corresponding to $2~ \Delta x$ (the Nyquist wavelength).
Figure \ref{fig-hpower} shows the two-dimensional power spectrum of
the total magnetic field and its vertical component at optical depth
unity in the case $\ReM = 100$: the power on small-scales close to the
resolution limit is much less than on the granulation scale. 
Hence in
the following, we concentrate on situations with $\ReM \le 400$.

\subsection{Kinematic models}

In the absence of non-linear effects, in case of a dynamo, one
expects a continuing exponential growth of ${\rm E}_{\rm M}$.
Figure~\ref{fig-emag} shows the result in terms of ${\rm E}_{\rm
M}(t)$ for five different models corresponding to varying \ReM:
\begin{figure}
\resizebox{\hsize}{!} {\includegraphics[]{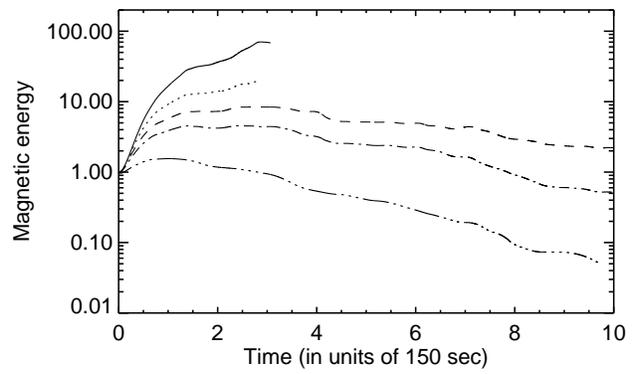}}
\caption{The total magnetic energy ${\rm E}_{\rm M}/{\rm E}_0$ as
a function of time in units of 150 seconds (the typical convective
turn-over time). Five models are presented, with different amounts
of diffusion: \ReM = 800 ({\bf full}), 400 ({\bf dotted}), 200
({\bf dashed}), 100 ({\bf dashed dotted}), and 20
 ({\bf dashed triple dotted curve}).}
\label{fig-emag}
\end{figure}
most of the models in fact are dominated by decaying modes, with
negative growth rates $\gamma_{\eta} < 0$. In particular, the case
with \ReM = 20 (high diffusion) is clearly an example of an
anti-dynamo: the diffusion works faster than the flows can sweep
up the field and concentrate it in the inter-granular downdraft
lanes, and the dominant magnetic mode is a decaying one.
Increasing \ReM\ decreases $\gamma_{\eta}$ (numerically), so that
for \ReM = 200 the decay time is 40 time longer than at \ReM = 20.
In terms of providing dynamo action, the most promising cases are
those with lower diffusion and \ReM $>$ 300--400: at \ReM = 400 a
growing mode seems to be dominating. One of the models in Fig.\
\ref{fig-emag} had \ReM = 800, i.e.\ it produced magnetic
structures too small to be resolved, and therefore magnetic energy
is lost from the computational domain and ${\rm E}_{\rm M}$ seems
to saturate because of enhanced numerical diffusion.

In the high diffusion case (e.g.\ \ReM = 20) the magnetic field varies
smoothly across the domain, and its power peaks on scales
approximately $\ell \approx 50\pun{km}$ when considering
$B_\mathrm{z}$ (see Fig~\ref{fig-convbz}).  At higher \ReM\ more small
scales are generated mostly around the downdraft lanes.

\subsection{Quenched models}

\begin{figure}
\resizebox{\hsize}{!}
{\includegraphics[]{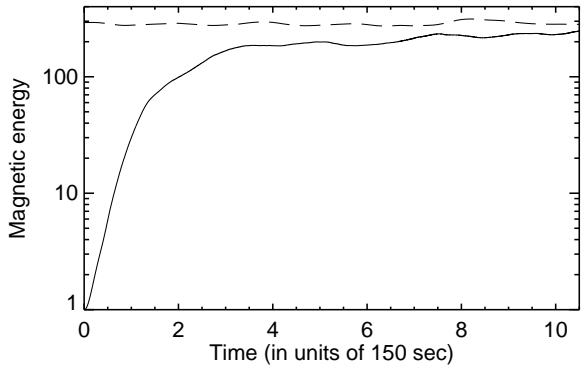}}
\caption{The total magnetic energy ${\rm E}_{\rm M}/{\rm E}_0$ as a
 function of time for a model with \ReM = 400 and a quenched
 velocity field ({\bf full}). Also shown is
 the total available kinetic energy ({\bf dashed curve}).}
\label{fig-dynamo}
\end{figure}

To examine the structure of the magnetic field in the
high \ReM\ cases with dynamo action, we computed a few models including
the quenching mentioned above:
quenching the velocity field only affects those models that are
dominated by growing modes ($\gamma_{\eta} > 0$). In case of a low
\ReM, the results remain unchanged since the field never becomes
comparable to ${\rm B}_{\rm eq}$. For \ReM = 200 the dominant mode
is still decaying. At \ReM = 400 the field is amplified
until it approaches equipartition and the quenching causes ${\rm
E}_{\rm M}$ to saturate; see Fig.\ \ref{fig-dynamo}. At the time
of saturation there is a lot of power at small scales close to the
resolution limit, but the main power is concentrated at larger
scales around 100 km. The saturation level is artificially fixed
by our choice of quenching function Eq.~(\ref{quench.eq}), the
value of $\alpha$, and our choice of a
constant value of $e_{\rm K}$. Hence our simulation results in no
direct information on the amplitude of the magnetic field, only on
the geometry that one may expect in a non-linear simulation.

\begin{figure}
\resizebox{\hsize}{!}
{\includegraphics[]{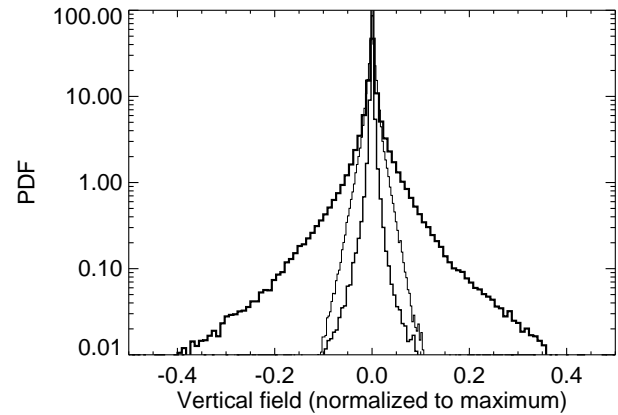}}
\caption{The probability distribution function (PDF) of the vertical field
component around optical depth unity. Lines show the distribution in
unquenched models for magnetic Reynolds numbers
\mbox{$\ReM=20,400,200$} ``outside-in'' of the hat shaped PDFs. Note
the non-monotonic behaviour with \ReM.}
\label{fig-pdf}
\end{figure}

Figure \ref{fig-pdf} shows the PDFs of $B_\mathrm{z}$ for three 
different values of
\ReM: the broad distribution of field strengths for low \ReM\ should
be seen in contrast to the narrow one for high \ReM. At any time
the strongest part of the field occupies only a very small part of the
volume no matter what the level of diffusion is, but for low \ReM\ the
preference for strong fields is less pronounced; the field is less
intermittent. We attribute the non-monotonic behavior of the models
displayed in Fig.~\ref{fig-pdf} to the model with
$\ReM=400$ showing dynamo action while the others do not.

\section{Conclusions}

\begin{itemize}
\item 
The geometry of small-scale (i.e. locally generated or
modulated) magnetic fields looks similar to the situation for the Sun
(i.e.  mostly located in the intergranular lanes). The basic reason is
that M-dwarf granulation is qualitatively similar to solar
granulation.

\item 
There are differences due to potentially quite different magnetic
diffusivities (and values of \ReM).

\item Depending on \ReM\, we get or do not get local dynamo action.

\item 
Since we are making a kinematic study, the absolute field strength is
undefined. Scaling from the Sun would suggest flux tubes with up to
factor 10 higher fields (assuming equipartition with the gas
pressure), or slightly sub-solar (within a factor 2) fields (assuming
equipartition with the kinetic energy density).

\end{itemize}

\acknowledgements
SBFD was supported through an EC-TMR grant to the European Solar
Magnetometry Network. HGL acknowledges financial support of the Walter
Gyllenberg Foundation, and the Swedish Vetenskapsr{\aa}det.


\begin{thebibliography}{}
\bibitem[2002]{Archontis+ea02} Archontis, V.D., Dorch, S.B.F., Nordlund, 
 {\AA}.: 2002, submitted to A\&A, preprint astro-ph/0204208 
\bibitem[2002]{Berger02} Berger, E.: 2002, preprint astro-ph/0111317
\bibitem[2000]{Dorch2000} Dorch, S.B.F.: 2000, Physica Scripta 61, 717 
\bibitem[1997]{Galsgaard+Nordlund97} Galsgaard, K., Nordlund, {\AA}.: 1997, 
 J.\ Geoph.\ Res.\ 102, 219 
\bibitem[2000]{Gizis+ea00} Gizis, J.E., Monet, D.G., Reid, N.I., 
Kirkpatrick, J.D., Libert, J., Williams, R.J.: 2000, AJ\,120, 1085
\bibitem[1996]{Johnskrullvalenti96} Johns-Krull, C.M., Valenti, J.A.:1996, AJ\,459, L95
\bibitem[2001]{KPVJ01} Kochukhov, O.P., Piskunov, N.E., Valenti, J.A., Johns-Krull, C.M.: 2001,
in: {\it The $\mathit{11}^\mathit{th}$ Workshop on Cool Stars, Stellar Systems and the Sun\/},
ASP~Conference Series, Vol.~223, 
eds. R.J.~Garc\'{\i}a L\'{o}pez, R.~Rebolo, M.R.~Zapatero Osorio, p.~985
\bibitem[1957]{Kopecky57} Kopeck{\'y}, M.: 1957, Bull. Astron. Inst. Czech~8, 71
\bibitem[2002]{LAH02} Ludwig, H.-G., Allard, F., Hauschildt, P.H.: 2002, submitted to A\&A
\bibitem[1999]{MM99} Meyer, F., Meyer-Hofmeister, E., 1999, A\&A\,341, L23
\bibitem[2002]{MBSAC02} Mohanty, S., Basri, G., Shu, F., Allard, F., Chabrier, G., 2002,
ApJ\,571, 469
\bibitem[1998]{SN98} Stein, R.F., Nordlund, {\AA}.: 1998, ApJ\,499, 914
\bibitem[1999]{SGBBN99} Stein, R.F., Georgobiani, D., Bercik, D.J., Brandenburg, A., 
Nordlund, {\AA}.: 1999, in: {\it Theory and Tests of Convection in Stellar Structure\/}, 
ASP Conference Series, Vol.~173, 
eds. Gim{\'e}nez, {\'A}, Guinan, E.F, Montesinos, B., p.~193
\bibitem[1989]{Stix89} Stix, M.: 1989, {\it The Sun: An Introduction\/}, Springer
\end{thebibliography}
\end{document}